\documentclass[openacc]{rstransa}

\newcommand{\dd}{\text{d}}

\newcommand{\DD}{{\cal D}}

\newcommand{\p}{\partial}
\newcommand{\vx}{\mathbf{x}}
\newcommand{\vxp}{\vx'}
\newcommand{\vq}{\mathbf{q}}
\newcommand{\vqp}{\vq'}

\newcommand{\vp}{\mathbf{p}}
\newcommand{\vk}{\mathbf{k}}
\newcommand{\vv}{\mathbf{v}}
\newcommand{\vbv}{\bar{\mathbf{v}}}
\newcommand{\vu}{\mathbf{u}}
\newcommand{\vbu}{\bar{\mathbf{u}}}
\newcommand{\vf}{\mathbf{f}}
\newcommand{\vj}{\mathbf{j}}
\newcommand{\vbj}{\bar{\mathbf{j}}}
\newcommand{\vnabla}{\mathbf{\nabla}}

\newcommand{\bla}{\big\langle}
\newcommand{\bra}{\big\rangle}

\usepackage[dvipsnames]{xcolor}

\titlehead{Research}

\begin{document}

%%%% Article title to be placed here
\title{Emergent dynamical scaling in the inviscid limit of 3D  stochastic  Navier-Stokes equation with thermal noise.}

\author{%%%% Author details
Liubov Gosteva$^{1}$, Marc Brachet$^{2}$ and L\'eonie Canet$^{1}$}

%%%%%%%%% Insert author address here
\address{$^{1}$Universit\'e Grenoble Alpes, CNRS, LPMMC, 38000 Grenoble, France\\
$^{2}$Laboratoire de Physique de l'Ecole Normale Sup\'erieure, CNRS, PSL Research University,
Sorbonne Universit\'e, Universit\'e Paris Cit\'e, F-75005 Paris, France}

%%%% Subject entries to be placed here %%%%
%\subject{xxxxx, xxxxx, xxxx}

%%%% Keyword entries to be placed here %%%%

%%%% Insert corresponding author and its email address}
\corres{L\'eonie Canet\\
\email{leonie.canet@lpmmc.cnrs.fr}}

%%%% Abstract text to be placed here %%%%%%%%%%%%
\begin{abstract}
 In this work, we investigate the Navier-Stokes equation in the presence of thermal noise, both at finite viscosity (revisiting the seminal work by Forster-Nelson-Stephen) and in the inviscid limit, which has not yet been explored. We determine the space-time velocity correlations in this dynamics,  using functional renormalisation group and direct numerical simulations.  While spectrally truncated three-dimensional  Euler flows reach a stationary equilibrium state, they exhibit non-trivial temporal correlations. We show that these non-trivial correlations persist for small but finite viscosity, yielding an emergent $\tau\sim k^{-1}$ dynamical scaling, where $\tau$ is the decorrelation time.
 We characterise the crossover  from the scaling $\tau\sim 1/(\nu k^2)$, expected at large viscosity, to the scaling $\tau\sim 1/(u_{\rm rms}k)$ found  in the inviscid limit. 
 % We find that, although the static properties of spectrally truncated %Euler flows radically differ from their hydrodynamical limit, they share a similar emergent $k^{-1}$ dynamical scaling.
\end{abstract}
%%%%%%%%%%%%%%%%%%%%%%%%%%%

%%%%%%%%%% Insert the texts which can accomdate on firstpage in the tag "fmtext" %%%%%

\begin{fmtext}

\section{Introduction}

Deriving the statistical theory of turbulence from the fundamental hydrodynamical equations has remained an unsolved challenge, despite intensive efforts since the pioneering works of Kolmogorov in 1941.
To gain some understanding, models such as the three-dimensional (3D) Euler equation or the one-dimensional (1D) Burgers equation, have played a pivotal role.
A noticeable feature of these models is that, while the continuous equations may generate  singularities in finite time~\cite{Bec2007,Eyink2008,Bustamante2012,Cordoba2023}, their  \end{fmtext}

%%%%%%%%%%%%%%% End of first page %%%%%%%%%%%%%%%%%%%%%

\maketitle

\noindent spectrally (Galerkin)-truncated versions lead to thermalisation to an equilibrium state~\cite{Lee1952,Kraichnan73,Majda2000,Ray2011,Murugan2020,Murugan2023,Cichowlas2005,Ray2015,Cameron2017}.
The static properties of the continuous and spectrally-truncated models thus radically differ.  For the 3D Euler flow, the continuous equation is expected to yield a turbulent non-equilibrium state with Kolmogorov kinetic
energy spectrum $E(k)\sim k^{-5/3}$, whereas the spectrally-truncated equation leads to the equilibrium state characterised by an equipartition energy spectrum $E(k)\sim k^{2}$. Note that the equipartition spectrum can also be observed in turbulent Navier-Stokes flows in the presence of thermal noise at sufficiently small scales in the dissipative range~\cite{Bandak2021,Bell2022,Betchov1957}.

  The thermalisation process to equilibrium has been thoroughly characterised, and was shown to involve localised oscillatory structures called {\it tygers}, both for the 1D  Burgers and for the 3D Euler spectrally-truncated  equations~\cite{Majda2000,Ray2011,Murugan2020,Murugan2023}.
  The emergence of these tygers deeply imprints the  dynamics, yielding a relaxation time  $\tau(k)\sim k^{-z}$ with $z=1$, both in the 1D Burgers~\cite{Majda2000} and in the 3D Euler~\cite{Cichowlas2005,Cameron2017} equations. Interestingly, when a thermal noise is added to  the deterministic 1D Burgers equation, the latter maps to the Kardar-Parisi-Zhang (KPZ) equation~\cite{Kardar86}, a celebrated Langevin-type model in statistical physics which features non-equilibrium scaling phenomena~\cite{Halpin-Healy95,Takeuchi18}. It was shown in direct numerical simulations that the stochastic 1D Burgers--KPZ equation also exhibits a decorrelation time scaling $\tau(k)\sim k^{-1}$ in the inviscid limit, as its deterministic counterpart~\cite{Brachet2022,Rodriguez2022}.

  This finding weaves a deep connection between statistical physics and deterministic hydrodynamics. It turns out that this link has not been explored yet for the spectrally-truncated 3D Euler equation. The aim of this paper is to bridge this gap, by investigating the stochastic Navier-Stokes equation in the presence of a thermal noise, and uncovering what happens when taking the inviscid limit of this model.
  The stochastic version of the 3D Navier-Stokes equation, including thermal noise, was  introduced in the seminal paper by Forster, Nelson, Stephen (FNS)~\cite{Forster76,Forster77}, and named model A.  The authors showed that this equation also leads to  relaxation to an equilibrium state, and thus model A  corresponds to a Langevin model for a fluid near thermal equilibrium. Let us notice that such studies of near-equilibrium (compressible) fluids date back to the early works by Landau and Placzek~\cite{Landau1934,Cummins1966} and have important applications. 
  However,  the  stochastic Euler dynamics was not discussed by FNS, as their approach relies on perturbative dynamical RG, which does not allow one to access the non-perturbative inviscid limit.

Beyond the theoretical interest, let us emphasise the physical motivation to revisit  model A. Let us remind that the hydrodynamical description is an effective description valid down to some length scale, which we denote $\Lambda^{-1}$. Hence, one has to account for fluctuations coming from the molecular dynamics below this scale. Adding a thermal noise, thereby  yielding the  stochastic Navier-Stokes equation, is therefore physically relevant (see for example Refs.~\cite{Bandak2021, Donev2014} and references within). The work~\cite{Bandak2021} highlights the appearance of the thermal noise as an observable effect, at lengthscales where the fluctuating hydrodynamics is still valid (see Eq.~(2.14) in~\cite{Bandak2021} and the discussion after). This was confirmed by numerical simulations~\cite{Bell2022}. In fact, the observation of  an equipartition spectrum $E(k)\propto k^2$ at lengthscales comparable to the Kolmogorov scale  was first reported in incompressible fluid turbulence experiments~\cite{Betchov1957}. In this (dissipative) range, the turbulent energy spectrum decays exponentially~\cite{Frisch1995}, such that the thermal fluctuations  dominate over the turbulent ones. This provides a further motivation to include the thermal noise in the Navier-Stokes equation.

In this work, we thus focus on model A, and study its yet unexplored inviscid limit,  using both numerical simulations and functional renormalisation group (FRG).
  For the numerical part, we perform direct numerical simulations of  model A of FNS, {\it ie} we integrate the spectrally-truncated stochastic 3D Navier-Stokes equation, in a Taylor-Green configuration, which allows us to access large system sizes. We compute the spatio-temporal two-point correlation function for different microscopic viscosities to determine the behaviour of the half decay time $\tau_{1/2}(k)$. We find that it also displays the $z=1$ scaling in the inviscid limit, as for the deterministic 3D Euler dynamics.

  To present the FRG part, let us first summarise previous achievements on which this work builds.
  For the stochastic 1D Burgers--KPZ equation, the origin of the $z=1$ scaling has been recently elucidated using FRG~\cite{Fontaine2023InvBurgers}.
   Indeed, it was shown to be controlled by a new fixed point of the Burgers--KPZ equation termed inviscid Burgers (IB) fixed point, and characterised by a critical dynamical exponent $z=1$, differing from the well-known KPZ value $z=3/2$. The FRG is a field-theoretical method based on the continuous equations. This salient scaling of the decorrelation time is not affected by the spectral truncation, although the static properties are drastically altered.
   Let us notice that for the 3D Navier-Stokes equation in the turbulent state, the emergence of a $z=1$ scaling is well-known phenomenologically, and interpreted as a consequence of  random sweeping effect~\cite{Tennekes75,Chen89,Nelkin90,Sanada92,Favier2010,Chevillard2012,diLeoni2015,Gorbunova2021}. For the forced Navier-Stokes, a rigorous derivation of this $z=1$ was provided using FRG~\cite{Canet2017,Tarpin2018,Canet2022}. Let us notice that it has been  proposed in~\cite{Lvov1987}, and used in several subsequent works, to move from the Eulerian to a quasi-Lagrangian frame in order to get rid of the sweeping effect, which   hinders the perturbative approaches~\cite{Chen89}. However, within the FRG formalism, which is non-perturbative, the sweeping effect is not problematic, and does not prevent from correctly describing the subleading straining turbulent dynamics and the Kolmogorov spectrum~\cite{Canet2022}. Thus, it is not necessary in this formalism to resort to a quasi-Lagrangian description, which is more complicated than the  Eulerian one.

Here, we use FRG to revisit  model A of FNS, focusing in particular on its inviscid limit. We show that, beside the attractive (IR) fixed point identified by FNS, there exists another repulsive (UV) fixed point, associated with zero viscosity, which we call inviscid FNS (IFNS), in analogy with the IB fixed point for the stochastic Burgers--KPZ equation, and which is also characterised by a $z=1$ critical dynamical exponent.  Thus, for both the stochastic 1D Burgers and 3D Navier-Stokes equations in equilibrium conditions, a non-trivial behaviour emerges in the inviscid limit, which is distinct from the KPZ or FNS one. Moreover,  we show that this behaviour is also relevant in the viscous case, where the corresponding scaling is observable at large momentum, over a range that extends when the viscosity is further reduced.

The remainder of the paper is organised as follows. In Sec.~\ref{sec:SNS}, we introduce  model A of FNS, and review the main results obtained by FNS which are relevant for this work. In Sec.~\ref{sec:FRG}, we briefly present the FRG formalism, and derive the flow equations for model A. We establish the fixed point structure of model A, and show that the dynamical exponent at the IFNS fixed point is $z=1$, both through a numerical integration of the complete flow equations within some approximation, and from an analytic solution of exact asymptotic flow equations in the large momentum limit. In Sec.~\ref{sec:DNS}, we present the results of the direct numerical simulations of model A, and compare with the FRG results. Finally, we discuss the physical relevance of the IFNS scaling in  Sec.~\ref{sec:exp} before giving our conclusions  in Sec.~\ref{sec:conclusion}.

\section{Stochastic Navier-Stokes equation}
\label{sec:SNS}

\subsection{Model A of FNS}
We consider the stochastic Navier-Stokes equation for an incompressible fluid in dimension $d$
\begin{align}
\label{eq:SNS}
 \p_t \vv + \lambda (\vv\cdot\vnabla) \vv &= -\dfrac{1}{\rho} \vnabla\pi+\nu \nabla^2 \vv +\vf\,\\
 \vnabla \cdot \vv &= 0\,,
\end{align}
where $\vv(t,\vx)$ is the fluid velocity field, $\pi(t,\vx)$ the pressure field, $\nu$ the (microscopic) kinematic viscosity of the fluid, $\rho$ its density, and $\lambda$ a real parameter introduced for later purpose and  eventually set to unity. The field $\vf$ is a stochastic forcing with Gaussian distribution of zero mean and of two-point correlations given
in Fourier space by
\begin{equation}
\label{eq:forcing}
\bla f_\alpha(\omega,\vq) f_\beta(\omega',\vqp) \bra =  2 \bar{D}(|\vq|) P^\perp_{\alpha\beta}(\vq) (2\pi)^{d+1}\delta(\omega+\omega')\delta^d(\vq+\vqp)\,,
\end{equation}
where $P^\perp_{\alpha\beta}(\vq) = \delta_{\alpha\beta} -\frac{q_\alpha q_\beta}{q^2}$ is the transverse  (Leray) projector. The Fourier conventions used throughout this paper are
\begin{align}
 g(\omega,\vq) &= {\cal F}\big[g(t,\vx)\big]= \int_{t,\vx} \,g(t,\vx)\, e^{i(\omega t -\vp\cdot\vx)}\,, \nonumber\\
 g(t,\vx) &= {\cal F}^{-1}\big[g(\omega,\vq)\big]=\int_{\omega,\vq} \,g(\omega,\vq)\, e^{-i(\omega t -\vp\cdot\vx)} \label{eq:Fourier}\,,
\end{align}
with the short-hand notations
\begin{equation}
  \int_{t,\vx}\equiv \int \dd^d \vx \dd t \,,\quad \int_{\omega,\vq} \equiv \int \dfrac{\dd^d \vq}{(2\pi)^d}\dfrac{\dd \omega}{2\pi}\,.
\end{equation}
We consider an infinite volume fluid, which we assume to be homogeneous and isotropic, and we focus on the stationary state.
The system is thus translationally invariant in space and time, and as a consequence the total frequency and momentum are conserved. Thus, the Fourier transform of a generic $n$-point function depends on only $n-1$ frequencies and momenta since the last ones are fixed by conservation. We denote with an overbar the Fourier transform without the factor $(2\pi)^{d+1}\delta\left(\sum_{k=1}^n\omega_k\right)\delta^d\left(\sum_{k=1}^n\vq_k\right)$ stemming from the conservation laws.

The choice $\bar{D}(q)=D q^2$ models a thermal noise, and the set of equations \eqref{eq:SNS}--\eqref{eq:forcing} identifies in this case with model A of FNS~\cite{Forster76,Forster77}.
It corresponds to a Langevin dynamics for a fluid near thermal equilibrium.  In particular, it satisfies a fluctuation-dissipation theorem which fixes the amplitude of the noise to $D  = \frac{\nu}{\rho} k_B T$ where $T$ is the temperature of the fluid and $k_B$ the Boltzmann constant. This system relaxes at long time to a stationary equilibrium state,  with equipartition of energy, yielding a kinetic energy spectrum $\bar{E}(k)\sim k^{d-1}$. This stochastic dynamics is invariant under the time-reversal symmetry defined by the tranformation~\eqref{eq:defTRS} (see next section). For this reason, we refer to the stochastic Navier-Stokes with thermal noise (model A) as the `equilibrium' dynamics, as it leads to an equilibrium stationary state (fluid at rest), while the Navier-Stokes equation with forcing at large scale is referred to as `non-equilibrium' since it leads to a stationary turbulent state. The forcing at large scales indeed breaks the symmetry~\eqref{eq:defTRS} and the fluctuation-dissipation relation no longer holds in the turbulent state \footnote{Note that in this work, equilibrium does not refer to the Euler dynamics and non-equilibrium to the  Navier-Stokes dynamics, which is another commonly used definition.}.

In this state, the velocity has a Gaussian distribution and its two-point correlation function takes the scaling form
\begin{equation}
\label{eq:scaling}
 \bar{G}^{(2,0)}_{\alpha\beta}(\omega,\vq) \equiv {\cal F}\big[\bla v_\alpha(t,\vx) v_\beta(t',\vxp)\bra_c \big] = P^\perp_{\alpha\beta}(\vq) q^{-2} \Phi(\omega/q^z)
\end{equation}
where the subscript $c$ stands for connected, and where $z$ is the dynamical critical exponent.
For model A,  one simply has $z=2$, characteristic of a diffusive behaviour,
with the universal scaling function $\Phi$ given by
 \begin{equation}
 \label{eq:phi}
 \Phi(x) = \chi\nu_{\rm eff}/(x^2+\nu_{\rm eff}^2)\,, \qquad \hbox{with}\qquad \chi = D /\nu= k_B T/\rho\,,
 \end{equation}
 and denoting $\nu_{\rm eff}$  the effective (measured) viscosity~\cite{Forster77}.

The relaxation process towards the equilibrium state gives rise to long-time tails, a phenomenon early  discussed in the litterature~\cite{Pomeau1975,Forster77}. Casting the correlation function in the form
\begin{equation}
\label{eq:nuR}
  \bar{G}^{(2,0)}_{\alpha\beta}(\omega,\vq) = 2 P^\perp_{\alpha\beta}(\vq)\Re\left[\dfrac{\chi}{-i \omega + q^2 \nu_R(\omega,q)}\right]
\end{equation}
 to define the renormalised viscosity $\nu_R(\omega,q)$, the long-time tail corresponds to a slow decay  as $\nu_R(t,q=0)\sim t^{-d/2}$ for $d>2$, with logarithmic corrections in $d=2$.

The constitutive equations of model A can be rescaled, such that there remains a single
  parameter $\bar\lambda = \lambda D^{1/2}/{\nu}^{3/2}$.
As emphasised by FNS~\cite{Forster77}, the equilibrium state corresponds to a vanishing
 renormalised $\bar\lambda$. This is why the correlations~\eqref{eq:scaling}--\eqref{eq:phi} are similar to results from  linearised hydrodynamics. Moreover,  the near equilibrium dynamics can be captured using perturbation theory, which is organised as a power series in
$\bar\lambda$.  However, this approach is not suitable to study fully developed turbulence, which corresponds to a non-trivial steady-state with a finite $\bar\lambda$ and non-Gaussian fluctuations.
 The perturbation theory in $\bar\lambda$  obviously also fails to study, even in the equilibrium state,  the inviscid limit $\nu\to 0$, as it corresponds to $\bar\lambda\to\infty$. For this, one has to resort to non-pertubative methods, such as the FRG which we explain in Sec.~\ref{sec:FRG}.

\subsection{Action and symmetries for model A}
\label{sec:action}

In order to apply  renormalisation group techniques, one can cast the stochastic dynamical equation into a path integral formulation, following the  Martin-Siggia-Rose-Janssen-de Dominicis  formalism~\cite{Martin73,Janssen76,Dominicis76}.
One obtains from the stochastic Navier-Stokes equation with thermal noise~\eqref{eq:SNS}--\eqref{eq:forcing} the following generating functional and action~\cite{Canet2022}
\begin{align}
 {\cal Z}[\vj,\vbj,K,\bar{K}] &= \int \DD \vv\DD \vbv \DD\pi \DD\bar{\pi} \,e^{-{\cal S}[\vv,\vbv,\pi,\bar{\pi}] + \int_{t,\vx} \big\{ \vj \cdot\vv + \vbj\cdot\vbv + K \pi +\bar{K}\bar{\pi} \big\}}\label{eq:Z}\\
 {\cal S}[\vv,\vbv,\pi,\bar{\pi}]  &= \int_{t,\vx} \Bigg\{ \bar{v}_\alpha \Big(\p_t v_\alpha +\lambda v_\beta \p_\beta v_\alpha +\dfrac{1}{\rho} \p_\alpha \pi -\nu \p^2 v_\alpha \Big) +\bar{\pi}\p_\gamma v_\gamma -D  \big(\p_\alpha \bar{v}_\beta\big)^2 \Bigg\}\, \label{eq:S},
\end{align}
where $\vbv$  and $\bar{\pi}$ are the response velocity and response pressure fields, and $\vj,\vbj,K,\bar{K}$ are the sources linearly coupled to the fields.
 The functional  ${\cal W} = \ln{\cal Z}$ is the generating functional of connected correlation functions -- which are  the equivalent, for fluctuating fields instead of random variables, to cumulants. Indeed, the connected velocity correlation and response functions  are obtained as
 \begin{equation}\label{eq:Wmn}
    G^{(m,n)}_{\alpha_1...\alpha_{m+n}}(t_1,\vx_1;\hdots;t_{m+n},\vx_{m+n}) \equiv
   \dfrac{\delta^{m+n}{\cal W} }
    {\delta j_{\alpha_1}(t_1,\vx_1)  \hdots
    \delta\bar{j}_{\alpha_{m+n}}(t_{m+n},\vx_{m+n})}
    \Big|_{\vj, \vbj=0}\, ,
\end{equation}
where the $m$ first derivatives are with respect to $\vj$ sources and the $n$ last with respect to $\vbj$ sources.

Another important generating functional is the Legendre transform of ${\cal W}$, which is called the effective action $\Gamma$ and  is defined by
\begin{equation}
\label{eq:Legendre}
    \Gamma[\Psi]  =  \underset{\mathcal{J}}{\sup} \left[-\mathcal{W}[\mathcal{J}]+\int_{t,\vx}\mathcal{J}_i \Psi_i\right]\,,\qquad \hbox{where}\quad
 \Psi_i\equiv\langle\Phi_i\rangle_{\cal J}=\dfrac{\delta {\cal W}}{\delta {\cal J}_i}\,,
\end{equation}
with $\Phi=(\vv,\vbv,\pi,\bar{\pi})$ denoting the multiplet of fields, $\mathcal{J}=(\vj,\bar{\vj},K,\bar{K})$ the multiplet of  associated sources, and $\Psi=(\vu,\vbu,\Pi,\bar{\Pi})$
  the multiplet of expectation values of the fields in the presence of the sources ${\cal J}$.
 One can obtain from $\Gamma$ the vertex functions, defined as
 \begin{equation}\label{eq:Gammamn}
    \Gamma^{(m,n)}_{\alpha_1...\alpha_{m+n}}(t_1,\vx_1;\hdots;t_{m+n},\vx_{m+n}) \equiv
   \dfrac{\delta^{m+n} \Gamma }
    {\delta u_{\alpha_1}(t_1,\vx_1)  \hdots
    \delta\bar{u}_{\alpha_{m+n}}(t_{m+n},\vx_{m+n})}
    \Big|_{\vu, \vbu=0}\, ,
\end{equation}
where again the $m$ first derivatives are with respect to $\vu$ fields and the $n$ last with respect to $\bar{\vu}$ fields.
The knowledge of the set of $G^{(m,n)}$, or equivalently of the set of $\Gamma^{(m,n)}$, provides the statistical theory for the model under study.

Let us now briefly explain how the symmetries lead to exact relations between these functions, which are called  Ward identities, and which can be very efficiently exploited within the FRG formalism. In fact, it turns out to be particularly useful to consider extended symmetries, which correspond to continuous infinitesimal transformations of coordinates and fields which do not leave the action strictly invariant, but yields a variation at most linear in the fields. One can indeed derive from such extended symmetries exact Ward identities, which deeply constrain the set of  $G^{(m,n)}$ (or equivalently the set of  $\Gamma^{(m,n)}$)~\cite{Canet2015}.

This procedure has been expounded in details, in  {\it eg.} Refs.~\cite{Tarpin2018,Canet2022} for the Navier-Stokes equation, or in {\it eg.} Refs.~\cite{Canet2010,Fontaine2023InvBurgers} for the  Burgers--KPZ equation. Hence, we only provide here a summary, and refer the interested reader to these references for technical details.
The first noticeable extended symmetries are shifts of the pressure fields, $\pi(t,\vx)\to \pi(t,\vx)+ \epsilon(t,\vx)$, and $\bar\pi(t,\vx)\to \bar\pi(t,\vx)+ \bar\epsilon(t,\vx)$, with $\epsilon(t,\vx)$ and $\bar\epsilon(t,\vx)$ infinitesimal scalar functions of time and coordinates. The corresponding Ward identities impose that the variation of the effective action $\Gamma$ under these transformations is equal to that of the action ${\cal S}$, which entails that the whole pressure sector is not renormalised.
  A second extended symmetry is the time-dependent Galilean symmetry,
  which corresponds to the transformation
  \begin{align}
 \delta v_\alpha(t,\vx) =-\dot{\epsilon}_\alpha(t)+\epsilon_\beta(t) \p_\beta v_\alpha(t,\vx)\,, &\quad
 \delta \pi(t,\vx) =\epsilon_\beta(t) \p_\beta \pi(t,\vx)\,, \nonumber\\
 \delta \bar v_\alpha(t,\vx) =\epsilon_\beta(t) \p_\beta \bar
 v_\alpha(t,\vx)\,, &\quad \delta \bar \pi(t,\vx) = \epsilon_\beta(t) \p_\beta \bar \pi(t,\vx) \, ,
\label{eq:defGal}
\end{align}
with $\epsilon(t)$ an infinitesimal vectorial function of time.
The corresponding Ward identities exactly relate a vertex function $\Gamma^{(m+1,n)}$ with one vanishing  momentum among the first $m+1$ ({\it ie.} associated with a velocity)  to  lower order ones $\Gamma^{(m,n)}$.
We here simply give the  two lowest-order identities
\begin{align}
    &\bar\Gamma^{(2,1)}_{\alpha\alpha_1\alpha_2} (\omega,0;\omega_1,\vp_1) =
   - \frac{p_{1\alpha}}{\omega}\left[
        \bar\Gamma^{(1,1)}_{\alpha_1\alpha_2}(\omega_1+\omega,\vp_1)
        -
        \bar\Gamma^{(1,1)}_{\alpha_1\alpha_2}(\omega_1,\vp_1)
    \right],
   \nonumber \\
    &\bar\Gamma^{(1,2)}_{\alpha\alpha_1\alpha_2} (\omega,0;\omega_1,\vp_1) =
   - \frac{p_{1\alpha}}{\omega}\left[
        \bar\Gamma^{(0,2)}_{\alpha_1\alpha_2}(\omega_1+\omega,\vp_1)
        -
        \bar\Gamma^{(0,2)}_{\alpha_1\alpha_2}(\omega_1,\vp_1)
    \right]\,, \label{eq:wardGal}
\end{align}
general expressions for arbitrary $(m,n)$ can be found in Ref.~\cite{Canet2022}.

 The third important extended symmetry is the time-dependent response shift, corresponding to the transformation
\begin{equation}
 \delta \bar v_\alpha(t,\vx) =\bar \epsilon_\alpha(t)\,,\quad
 \delta \bar \pi(t,\vx)= v_\beta(t,\vx) \bar \epsilon_\beta(t) \, ,\label{eq:defShift}
\end{equation}
  with $\bar\epsilon(t)$ an infinitesimal vectorial function of time. The corresponding Ward identities exactly impose that a vertex function $\Gamma^{(m,n)}$ with one vanishing momentum among the last $n$ ({\it i.e} associated with a response velocity) vanishes, except the two lowest-order ones, which are fixed to
\begin{align}
&\bar \Gamma_{\alpha\beta}^{(1,1)}(\omega,\vp=\mathbf{0})=i\omega\delta_{\alpha\beta},\nonumber\\
&\bar\Gamma_{\alpha\beta\gamma}^{(2,1)}(\omega_1,\vp_1,\omega_2,-\vp_1) = i
p_1^\alpha \delta_{\beta\gamma} -ip_1^\beta \delta_{\alpha\gamma} \label{eq:wardShift}\, .
\end{align}

Finally, model A possesses an additional symmetry compared to Navier-Stokes equation (without thermal noise), which is a  time-reversal symmetry, corresponding to the  discrete transformation
\begin{align}
v_\alpha(t,\vx) \to -v_\alpha(-t,\vx) \,,\quad \pi(t,\vx) \to \pi(-t,\vx)\,, \nonumber\\
\bar{v}_\alpha(t,\vx) \to \bar{v}_\alpha(-t,\vx) -\dfrac{\nu}{D } v_\alpha(-t,\vx) \,,\quad \bar{\pi}(t,\vx) \to -\bar{\pi}(-t,\vx)\,.
\label{eq:defTRS}
\end{align}
This symmetry also yields constraints on the vertex functions, in particular
\begin{equation}
2\Re {\rm e}\Big[\bar{\Gamma}^{(1,1)}_{\alpha\beta}(\omega,\vp)\Big] = - \dfrac{\nu}{D }   \bar{\Gamma}^{(0,2)}_{\alpha\beta}(\omega,\vp)\, ,
\label{eq:wardTRS}
\end{equation}
which is a fluctuation-dissipation relation.
All these identities are exploited in the following to devise  suitable approximation schemes in the FRG formalism.

\section{Functional renormalisation group}
\label{sec:FRG}

\subsection{FRG formalism}

We briefly lay out the basis of the FRG formalism, referring the reader  to comprehensive introductions, such as~\cite{Berges2002,Kopietz2010,Delamotte2012,Dupuis2021}, for more details.  The FRG is based on the idea of Wilsonian RG. Instead of averaging over the fluctuating fields $\Phi$ in the path integral \eqref{eq:Z} at all (momentum) scales simultaneously, one only integrates them down to a certain RG scale $\kappa$. One then considers how this modified theory evolves under an infinitesimal change of the RG scale, which  gives rise to a deterministic differential equation, akin a dynamical system, where the role ot time is played by the RG scale, and the physical trajectories are replaced by trajectories in an abstract theory space. Such a trajectory  connects a given microscopic model to its effective theory at large distance and time. In analogy with  a fixed point in the dynamical flow  which corresponds to a stationary (time-independent) state,  a fixed point in the RG flow  corresponds to a scale-invariant state. 

From the technical point of view, the progressive averaging of fluctuations is achieved in FRG through the introduction of a scale-dependent `weight' $\exp(- \Delta \mathcal{S}_{\kappa})$ in the functional integral to suppress fluctuations below the RG scale $\kappa$:
\begin{equation}\label{Zkappa}
    \mathcal{Z}_{\kappa} = \int {\mathrm{D}} \Phi \exp \left(
        -\mathcal{S}[\Phi] - \Delta \mathcal{S}_{\kappa}[\Phi] + \int_{t,\vx} \mathcal{J}_i \Phi_i \right),
\end{equation}
where $\Delta \mathcal{S}_{\kappa}[\Phi] \equiv \frac{1}{2}\int_{t,\vx,\vx'} \Phi_i(t,\vx) \mathcal{R}_{\kappa,ij}(|\vx-\vx'|) \Phi_j(t,\vx')$. The kernel $\mathcal{R}_{\kappa}$ is called `regulator'. Its precise form is unimportant, as long as all its elements  satisfy the following set of properties (in  Fourier space): \\(i)
   $\mathcal{R}_{\kappa,ij}(\vp) \stackrel{\kappa \rightarrow \Lambda}{\sim} \Lambda^2$ such that  all fluctuations are frozen at the initial (microscopic) RG scale $\Lambda$;\\
  (ii) $\mathcal{R}_{\kappa,ij}(\vp) \stackrel{\kappa \rightarrow 0}{\longrightarrow} 0$ such that all fluctuations are averaged over at the final (macroscopic) RG scale;\\
   (iii) $\mathcal{R}_{\kappa,ij}(\vp) \stackrel{|\vp| \ll \kappa}{\sim} \kappa^2$ such that at a scale $\kappa$, the contribution of the `slow' modes is suppressed;\\
   (iv) $\mathcal{R}_{\kappa,ij}(\vp) \stackrel{|\vp| \gg \kappa}{\longrightarrow} 0$ such that at a scale $\kappa$, the `fast' modes are unaltered and averaged over,\\
   where we recall that $\Lambda^{-1}$ denotes the microscopic length scale at which the hydrodynamical model is defined.
In the presence of the regulator, the Legendre transform \eqref{eq:Legendre} is slightly modified
   to
\begin{equation}
    \Gamma_{\kappa}[\Psi] + \Delta\mathcal{S}_{\kappa}[\Psi] =  \underset{\mathcal{J}}{\sup} \left[-\mathcal{W}_{\kappa}[\mathcal{J}]+\int_{t,\vx}\mathcal{J}_i \Psi_i
    \right]\,,
\end{equation}
where ${\cal W}_\kappa =\ln{\cal Z}_\kappa$, and $\Gamma_\kappa$ is called the effective average action. In the large-$\kappa$ limit, $\Gamma_\kappa$ identifies with the microscopic action $\Gamma_\Lambda\equiv \mathcal{S}$  since all  fluctuations are frozen, while in the small-$\kappa$ limit, the full effective action  $\Gamma_0\equiv\Gamma$, encompassing all  fluctuations at all scales, is obtained~\cite{Delamotte2012,Dupuis2021}.  This provides $\Gamma_\kappa$ with a  physical interpretation as the effective theory at scale $\kappa$, interpolating from the microscopic theory  at $\kappa=\Lambda$ to the complete effective theory when $\kappa\to 0$. The very idea of the Wilsonian RG is thus realised through this formalism in a consistent and efficient way.
The evolution of $\Gamma_{\kappa}$ with the RG scale in between these two
limits is given by the exact Wetterich equation~\cite{Wetterich93,Ellwanger94,Morris94}
\begin{equation}\label{eq:Wetterich}
    \partial_{\kappa} \Gamma_{\kappa} =
    \frac{1}{2}\textrm{tr}\,\int_{\omega,\vq}
    \partial_{\kappa} \mathcal{R}_{\kappa}\,
    \mathcal{G}_{\kappa}\,,\qquad \hbox{with}\quad \mathcal{G}_{\kappa} \equiv \left(
    \Gamma_{\kappa}^{(2)} + \mathcal{R}_{\kappa}
\right)^{-1}\,,
\end{equation}
where the  trace means summation over all fields, $\Gamma_{\kappa}^{(2)}$ is the Hessian
of $\Gamma$, and
$\mathcal{G}_{\kappa}$  the full (functional) propagator of the theory.

The flow equation~\eqref{eq:Wetterich} is exact, but as it is a functional partial differential equation, it cannot be solved exactly in general. Several approximation schemes are available to study it, which are  non-perturbative and can be controlled and improved in a systematic way~\cite{Dupuis2021}. In the following, we present two complementary approaches to treat these equations for model A and investigate its inviscid limit. One of them consists in devising an Ansatz for $\Gamma_\kappa$, which allows one to project the flow onto a set of chosen renormalised functions, which are then numerically solved, as done in {\it eg.} Ref.~\cite{Canet2016}. This approach is particularly accurate to describe the small-momentum behaviour. It allows us to first qualitatively establish the fixed point structure of model A,  evidencing the existence of the IFNS fixed point.
The other approach consists in expanding the flow equation for the velocity correlation function at large momentum. It turns out that the leading order of this expansion can be closed exactly using the Ward identities stemming from the extended symmetries. Moreover, it can  be solved analytically, yielding the exact asymptotic expression of the correlation function at large momentum. This approach is thus appropriate to describe the large-momentum behaviour. It allows us to show in particular that $z=1$ exactly at the IFNS fixed point. Finally, to obtain a quantitative picture over the whole range of momenta, we consider an advanced numerical method which allows us to solve simulaneously the two sets (large- and small-momentum) of flow equations and compute the full correlation function and the associated half decay frequency $\omega_{1/2}(p)$.
 We start by presenting the large-momentum equations to establish the general result $z=1$, before turning to the small-momentum one to show the existence of the new IFNS fixed point.

\subsection{Exact closure at large momentum}

The exact FRG flow equations for any vertex function $\Gamma_\kappa^{(m,n)}$ can be obtained by taking $m$ functional derivatives of Eq.~\eqref{eq:Wetterich} with respect to  $\vu$ fields and $n$ with respect to  $\bar{\vu}$ fields. It turns out that they can be simplified in a controlled way in the limit of large momenta. This originates in the very structure of these flow equations, in particular in the presence of the regulator term ${\partial}_s \mathcal{R}_{\kappa}(\vq)$ in the loop integral. It implies that the integral over the internal momentum $\vq$ is cut to values
$|\vq|\lesssim \kappa$. Thus, in the limit where all the external momenta are large  $|\vp_i|\gg\kappa$ for all $i=1,\dots, m+n$,
 one has $|\vq|\ll |\vp_i|$. Because analyticity of the vertices is guaranteed  at all scales $\kappa$ by construction of the FRG formalism~\cite{Dupuis2021}, the vertices appearing in the flow equation can be Taylor expanded in powers of $\vq$, since $|\vq|$ is always compared to one of the $|\vp_i|$ and $|\vq|/|\vp_i|\to 0$ for all $i$. This expansion becomes exact in the limit where all the $\vp_i$ (and their partial sums) tend to infinity and in the vicinity of a fixed point~\cite{Blaizot2006,Blaizot2007,Benitez2012}. In practice, for turbulence, the  RG flow essentially stops when $\kappa \lesssim L^{-1}$ where $L$ is the integral scale, such that the large momentum condition becomes $|\vp_i|\gg L^{-1}$.

While the exactness of the large momentum expansion of the FRG flow always holds, an exceptional feature occurs in the context of turbulence, which stems from the existence of the extended symmetries highlighted in Sec.~2\ref{sec:action}. Indeed, the set of related Ward identities~\eqref{eq:wardGal},
 \eqref{eq:wardShift} (and the higher-order ones)  appears to fix all the vertices with one zero momentum (associated with a velocity or a response velocity) in terms of lower-order vertices. These are precisely the configurations which arise in the  flow equations expanded at large momenta ($|\vq|/|\vp_i|\to 0$), which can thus be closed exactly using the Ward identities. Moreover, the resulting closed flow equations  can be solved analytically in the vicinity of a fixed point. This exact closure was first achieved for the forced (non-equilibrium) Navier-Stokes equation in three, and then two dimensions, and for passively advected scalars~\cite{Canet2016,Canet2017,Tarpin2018,Tarpin2019,Pagani2021}. It was also demonstrated for the one-dimensional, and $d$-dimensional, Burgers--KPZ equation~\cite{Fontaine2023InvBurgers,Gosteva2024}. We refer the readers to these references or to the review~\cite{Canet2022} for the complete proof.
 
 The crucial point is that the Wetterich FRG flow equations for the $n$-point correlation functions are very generic, as well as their large momentum expansion. Moreover,  their exact closure only relies on the extended symmetries, and not on the details of the  specific hydrodynamical equation, in particular not on the precise form of the noise or forcing term, provided it has a Gaussian distribution, since this term does not enter the Ward identities~\eqref{eq:wardGal},
 \eqref{eq:wardShift}. In this respect, the large momentum equations of model A are the same as the ones obtained for the forced Navier-Stokes, except that there is only one independent two-point function because of the identity~\eqref{eq:wardTRS} stemming from time-reversal symmetry. We give it for the transverse correlation function $\bar C(\varpi, \vp) \equiv P_{\alpha\beta}^\perp(\vp) \bar G_{\alpha\beta}^{(2,0)}(\varpi,\vp)$ (the longitudinal part vanishes because of incompressibility)~\cite{Tarpin2018,Canet2022}
\begin{equation}
\p_s \bar C_{\kappa}(\varpi, \vp) =  \dfrac{d-1}{d} {p^2}
\int_{\omega} \dfrac{\bar C_{\kappa}(\omega+\varpi,  \vp) -
\bar C_{\kappa}(\varpi,  \vp)}{\omega^2}\,  {J}_\kappa(\omega)\label{eq:flowC}\,,\qquad
{J}_\kappa(\omega)\equiv \tilde{\p}_s \int_{\vq}\bar{C}_{\kappa}(\omega,  \vq)\,,
\end{equation}
with $\p_s\equiv \kappa \p_\kappa$,  and where $\tilde\p_s$ only acts on the scale dependence of the regulator. This equation can be solved at a fixed point.
Fourier transforming back to time delay, the solution reads~\cite{Tarpin2018,Canet2022}
\begin{equation}\label{eq:solutionC}
    \bar{C}(t,\vp) =  \bar{C}(0,\vp){\times}
    \begin{cases}
        \exp\left( - \mu_0 (pt)^2 \right),\quad t\ll \tau_c
        \\
        \exp\left( - \mu_{\infty} p^2 |t| \right),\quad t\gg \tau_c
    \end{cases}
\end{equation}
where $\mu_0$ and $\mu_{\infty}$ are non-universal constants (related to the fixed-point values of $J_\kappa(0)$ and $\int_\omega J_\kappa(\omega)$ respectively) and $\tau_c$ is a characteristic time scale. This expression shows that the correlation function takes at small time delays a scaling form similar to~\eqref{eq:scaling} with the scaling variable $x=pt$, and thus $z=1$, different from the FNS value $z=2$. Hence, even in the viscous case where the small momenta (large distance) converge to the scaling form~\eqref{eq:scaling} with $z=2$, the large momenta exhibit a different scaling when the viscosity is small enough. Moreover, the solution~\eqref{eq:solutionC} predicts a crossover at large time delays to yet another scaling regime, with an exponent $z=2$. However, this regime differs from the viscous diffusive regime since $\mu_{\infty}$ does not depend on $\nu$, the scaling form is different from \eqref{eq:phi}, and it emerges at large momentum (not at large distance). Let us note that the solution~\eqref{eq:solutionC} is an exact asymptotic result, but it does not give quantitative information about when ({\it i.e.} for how large momentum, and how small viscosity) it holds. Indeed, for actual numbers, one would have to integrate the whole flow (not only its asymptotic form at large $p$) to calculate $J_\kappa(\omega)$ for the precise model studied. This quantity is in particular sensitive to the  form of the noise or forcing. This issue will be investigated in Sec.~3\ref{sec:resFRG}.

As previously mentioned, the result~\eqref{eq:solutionC} is very general, it also holds for the large-momentum behaviour of the forced Navier-Stokes equation or passive scalar turbulence. It has been investigated in these contexts through direct numerical simulations.
While the small-time $z=1$ decay is easily observed, {\it eg.}~\cite{Favier2010,Gorbunova2021}, the large-time $z=2$ decay has remained elusive so far. Indeed,  this regime turns out to be hard to resolve in numerical simulations since the initial Gaussian decay is very fast and the large-time correlations fall below the noise level. This regime has been evidenced so far only for a generalisation of Kraichnan's model for passive scalars in Ref.~\cite{Gorbunova2021scalar}.
We show in Sec.~\ref{sec:DNS} that for model A, we manage to evidence both the small and the large time regimes in our direct numerical simulations, which is therefore an important result of the present work.

 %These different scaling regimes can  be evidenced in the behaviour of the half decay time $\tau_{1/2}(p)$, resp.  half decay frequency $\omega_{1/2}(p)$, defined for each momentum $p$ as the time, resp. frequency, at which the correlation function has decayed to half its value at $t=0$, resp. $\omega=0$ (see next sections).

\subsection{FRG flow equations for model A}
\label{sec:flowmodelA}

The asymptotic expression~\eqref{eq:solutionC} holds for large $p$. However, as previously mentioned, the constants  $\mu_0$, $\mu_\infty$ are related to the fixed-point values of $J_\kappa(\omega)$ in~\eqref{eq:flowC}~\cite{Tarpin2018}. They hence primarily depend on the small $p$ region, and their calculation requires the integration of the whole flow. An approximation which is accurate to describe the small momentum sector is to devise an ansatz for $\Gamma_\kappa$. Taking into account all the symmetries and associated non-renormalisation theorems, the appropriate ansatz reads~\cite{Canet2016}
\begin{equation}
  \Gamma_\kappa = \int_{t,\vx} \Bigg\{ \bar{v}_\alpha \Big(\p_t v_\alpha + \lambda v_\beta \p_\beta v_\alpha +  \frac{1}{\rho}\p_\alpha \pi - \nu_\kappa(\p_t,\vnabla) \p^2 v_\alpha \Big) +\bar{\pi}\p_\gamma v_\gamma -\chi \nu_\kappa(\p_t,\vnabla) \big(\p_\alpha \bar{v}_\beta\big)^2 \Bigg\}\, \label{eq:Gammak}\,,
\end{equation}
where  in Fourier space $\bar{\nu}_\kappa(\omega,\vp)$ is a whole function of $\omega$ and $\vp$, which identifies with the renormalised viscosity defined by~\eqref{eq:nuR}. 
Note that the same function $\bar{\nu}_\kappa$ appears for both the dissipation and the noise term, as a consequence of the time-reversal symmetry~\eqref{eq:wardTRS}. However, in contrast with perturbative calculations, the FRG allows one to compute the flow of this whole function (hence the F in FRG), and not only renormalised parameters as standard dynamical RG. The ansatz~\eqref{eq:Gammak} encompasses the most general two-point functions, while it neglects all higher-order vertices but $\Gamma_\kappa^{(2,1)}$ which is present at the bare level and is not renormalised. It was shown to yield accurate results for 3D and 2D turbulence~\cite{Canet2022}.

Projecting the exact Wetterich equation~\eqref{eq:Wetterich} onto the ansatz~\eqref{eq:Gammak}, one obtains the flow equation of $\bar{\nu}_\kappa(\omega,\vp)$ as
\begin{align}
\partial_s \bar\nu_\kappa(\omega,\vp)
 &=\frac{2\lambda^2}{(d-1)\vp^2}\int_{\varpi,\vq}  \Bigg\{
 \partial_s R_\kappa(\vq) \dfrac{ \varpi^2 - \tilde\nu_\kappa(\varpi,\vq)^2}{\big(\varpi^2+(\tilde\nu_\kappa(\varpi,\vq))^2\big)^2}\dfrac{\tilde\nu_\kappa(\omega+\varpi,\vp+\vq)}{(\omega+\varpi)^2+ (\tilde\nu_\kappa(\omega+\varpi,\vp+\vq))^2}
\nonumber\\
&\times\Big[\Big(2\vp^2 -\frac{(\vp \cdot (\vp+\vq))^2}{(\vp+\vq)^2}-\frac{(\vp \cdot \vq)^2}{\vq^2}\Big)(d-1)\nonumber\\
 &+2 \frac{1}{\vq^2 (\vp+\vq)^2}\Big(\vq^2-\frac{(\vp\cdot \vq)^2}{\vp^2}\Big)
 \Big(\vp^2 \vp\cdot \vq +2 (\vp\cdot \vq)^2-\vp^2 \vq^2\Big)\Big]\Bigg\}\,,
 \label{eq:flownu}
\end{align}
where $R_\kappa$ denotes the component of the regulator matrix ${\cal R}_\kappa$ defined as $R_\kappa\equiv {\cal R}_{\kappa,12} = {\cal R}_{\kappa,21} =-{\cal R}_{\kappa,22} /2$,
 and where $\tilde\nu_\kappa(\varpi,\vq) = \vq^2 \bar{\nu}_\kappa(\varpi,\vq) + R_\kappa(\vq)$.
The derivation of this equation can be found in  Ref.~\cite{Canet2016}, simply imposing at the end the constraints ensuing from time-reversal symmetry~\footnote{The flow equation~\eqref{eq:flownu} corresponds to Eq. (B10) of Ref.~\cite{Canet2016}, setting  $f_\perp^D(p)=\chi f_\perp^\nu(\vp)$ and $N_\kappa=\chi R_\kappa$ to preserve time-reversal symmetry,
 and restoring the frequency arguments $f_\perp^\nu(\vq)\to\tilde\nu_\kappa(\varpi,\vq)$,  $f_\perp^\nu(\vp+\vq)\to\tilde\nu_\kappa(\omega+\varpi,\vp+\vq)$. Note that a sign typo in the last line of (B10) has been corrected.}. Note that a further simplification is used for the numerical resolution of this equation, which consists in neglecting the frequency dependence of $\tilde\nu_\kappa$ within the integral, which amounts to the replacement $\tilde\nu(\varpi,\vq)\simeq \tilde\nu_\kappa(0,\vq)$ in the integrand.

\subsection{Results from FRG}
\label{sec:resFRG}

\begin{figure}[!h]
\includegraphics[width=6.5cm]{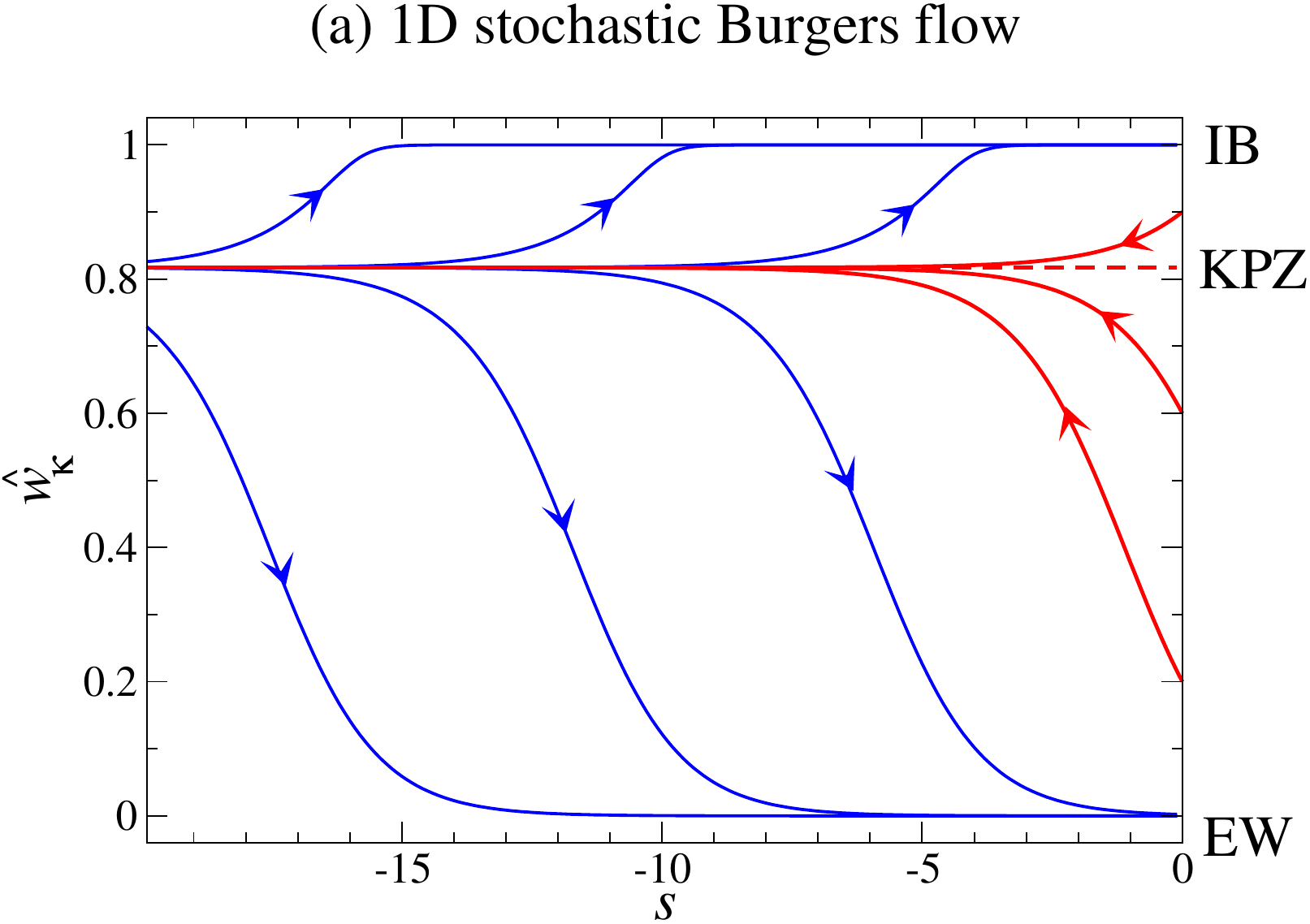}\includegraphics[width=6.5cm]{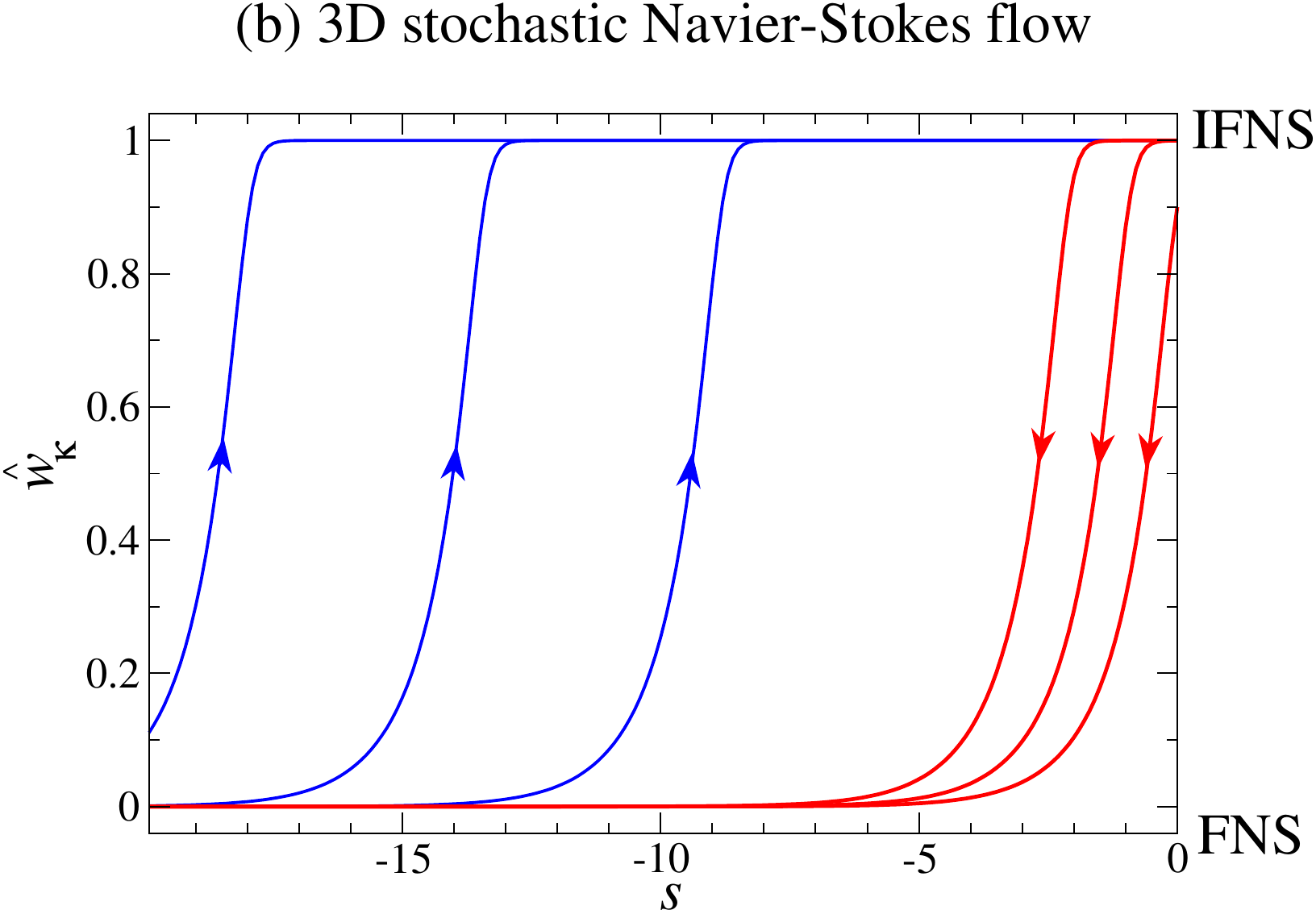}
\caption{FRG flows of the non-linear coupling $\hat w_\kappa$ for (a) the stochastic 1D  Burgers   and (b) the stochastic  3D Navier-Stokes equations: in red flows from $s=\ln(\kappa/\Lambda)=0$ to $s\to-\infty$ ($\kappa\to0$) towards the IR fixed points for different initial conditions $\hat w_\Lambda$; in blue flows from small initial $\kappa_0$ (large negative  $s\lesssim -20$) to $s\to0$ ($\kappa\to\Lambda$) towards the UV fixed points for different initial conditions $\hat w_{\kappa_0}$.}
\label{fig:flow}
\end{figure}

 We first elucidate the fixed point structure of the FNS equation within an even simpler approximation, which consists in considering only the (non-perturbative) flow of the non-linear dimensionless coupling parameter $\bar g_\kappa = \kappa^{d-2}\chi \lambda^2 /\nu_\kappa^2$, with $\nu_\kappa\equiv\bar{\nu}_\kappa(0,0)$. Since we are interested in the inviscid fixed point, corresponding to $\bar g_\kappa\to\infty$, we define $\hat w_\kappa = \bar g_\kappa/(1+\bar g_\kappa)$. Its flow is given by
 \begin{equation}
\label{eq:dsw}
\p_s \hat w_\kappa = \hat w_\kappa(1-\hat w_\kappa)\big(d-2 +2 \eta_\kappa \big)\,, \quad \hbox{with}\;\;\eta_\kappa = -\p_s \ln \nu_\kappa\, ,
\end{equation}
and the flow  $\p_s\nu_\kappa$ is obtained from~\eqref{eq:flownu} evaluated at zero external frequency $\omega=0$ and momentum $\vp=0$.
This equation can be straightforwardly integrated numerically. The result in shown in Fig.~\ref{fig:flow}, where it is compared with the equivalent flow for the stochastic 1D Burgers equation from Ref.~\cite{Fontaine2023InvBurgers}. From any initial condition $\hat w_\Lambda$ at  $\kappa=\Lambda$ ($s=\ln(\kappa/\Lambda)=0$), the flow reaches, as $\kappa\to 0$ ($s\to -\infty$), an IR fixed point corresponding to $\hat w^{\rm FNS}_*=0$. This is the attractive FNS fixed point. To evidence UV fixed points, one can run the flow backwards, starting from a very small initial $\kappa\equiv \kappa_0$ (very large negative initial $s$) to $\kappa\to\Lambda$ ($s\to 0$), such that the stability of the fixed points are exchanged. As illustrated on Fig.~\ref{fig:flow}(b), the backwards flow always reaches in the UV a new fixed point, corresponding to  $\hat w^{\rm IFNS}_*=1$,  which we call IFNS. This demonstrates the existence of this fixed point. In contrast, for the 1D stochastic Burgers equation, the RG flow always leads in the IR to a non-Gaussian fixed point with a finite $0<\hat w^{\rm KPZ}_*<1$ which is the KPZ fixed point. Running the flow backwards from  a very large negative value of $s$ (very small $\kappa_0$) to $s\to0$, the flow can reach  two different UV fixed points depending on the initial value of the coupling: the IB fixed point for $\hat w_{\kappa_0}> \hat w^{\rm KPZ}_*$, and the EW one for  $\hat w_{\kappa_0} < \hat w^{\rm KPZ}_*$, as illustrated on Fig.~\ref{fig:flow}(a).

\begin{figure}[!h]
\centering
\centering\includegraphics[width=12cm]{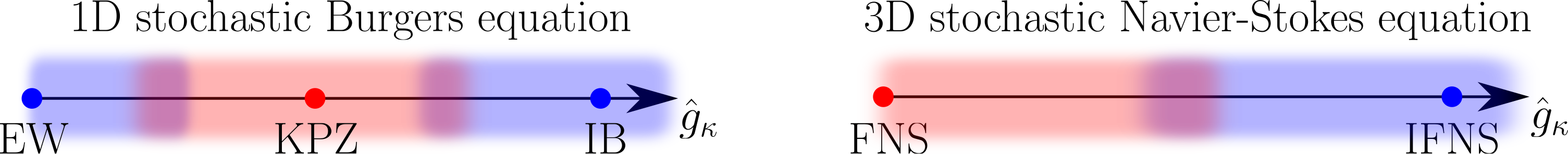}
\caption{Sketch of the fixed point structure of the stochastic 1D Burgers and  stochastic 3D Navier-Stokes equations in the presence of thermal noise. The red points represent attractive (IR) fixed points and the blue points repulsive (UV) ones. The red-shaded (resp. blue-shaded) area hence symbolises  the IR (resp. UV) modes controlled by the IR (resp. UV) fixed points.}
\label{fig:sketch}
\end{figure}

This unveils the whole fixed point structure of model A, which is sketched in Fig.~\ref{fig:sketch}, again in correspondence with the one of the 1D Burgers--KPZ equation. The FNS fixed point describes the behaviour at  large distance and time, {\it i.e.} the behaviour  of the IR modes (symbolised by the red-shaded area in the figure), of the viscous fluid at rest. The IFNS fixed point describes the inviscid fluid at equilibrium, but it also controls the large (UV) modes of a viscous fluid, the range of these modes (symbolised by the blue-shaded area in the figure) extending as the viscosity decreases.

\begin{figure}[!h]
%\centering\includegraphics[width=0.5\linewidth]{figure3.pdf}
\centering\includegraphics[width=6cm]{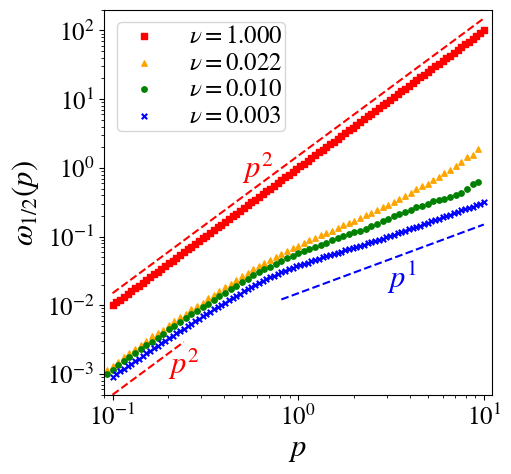}
\caption{Half decay frequency for different initial values of the viscosity. The IR modes show a diffusive FNS scaling $\omega_{1/2}\sim p^2$. The IFNS scaling, characterised by $\omega_{1/2}\sim p$, appears for UV modes when the viscosity decreases. In this figure, the momentum $p$ is measured in units of $\Lambda$  and the  viscosity $\nu$ in units of $(\Lambda\chi)^{1/2}$.}.
\label{fig:omega-half}
\end{figure}

In order to obtain more quantitative results, and in particular  the correlation function,
 we numerically solve the complete set of flow equations~\eqref{eq:flowC}, \eqref{eq:flownu}, using two complementary grids~\cite{Benitez2012}. A first grid is used for the small momentum sector, where the flow equation~\eqref{eq:flownu} is integrated using dimensionless variables with a similar numerical scheme as described in Refs.~\cite{Fontaine2023InvBurgers,Canet2016}. The large momentum flow equation~\eqref{eq:flowC} is integrated on a second grid using dimensionful variables, with the input at each scale $\kappa$ of $J_\kappa(\omega)$ computed on the dimensionless grid.
 The full numerical procedure is detailed in Ref.~\cite{Gosteva2025}.
 This scheme allows us to compute the correlation function $\bar{C}(\omega,\vp)$ on the whole momentum range.
 To evidence scaling regimes, we extract from it the half decay frequency defined such that $\bar{C}(\omega_{1/2}(p),\vp)=\bar{C}(0,\vp)/2$. The result is displayed on Fig.~\ref{fig:omega-half}.
Two markedly different scaling regimes are observable, thus unveiling the existence of the two  underlying fixed points. At small momenta, one finds $\omega_{1/2}\sim q^{z}$ with $z=2$ the FNS diffusive exponent. This region is hence controlled by the IR FNS fixed point as expected. For large viscosities ({\it eg} $\nu=1$), only this scaling is visible. For large momenta, when the viscosity decreases, one finds another scaling regime where $\omega_{1/2}\sim q^{z}$ with $z=1$. This regime  is all the more pronounced than the viscosity is small.

\section{Results from direct numerical simulations}
\label{sec:DNS}

\subsection{Numerical methods}
The stochastic Navier-Stokes equation with thermal noise~\eqref{eq:SNS}--\eqref{eq:forcing} are numerically solved using $2/3$ de-aliased Fourier spectral methods~\cite{Gottlieb1977}.
We thus use the discrete Fourier transform  $\vv(t,\vx)=\sum \hat{\vv}(t,\vk) e^{i \vk\cdot \vx}$
of a 3D spatially periodic velocity field.
Performing a Galerkin truncation on~\eqref{eq:SNS}--\eqref{eq:forcing}, {\it i.e.} ${\hat \vv}(t,\vk)=0$ for
$k_\alpha k_\alpha \ge k_{\rm max}^2$ (where $k_{\rm max}$ is $1/3$ of the resolution),
generates the following finite system of ordinary differential equations for the complex variables ${\hat \vv}(t,\vk)$ ($\vk$ being a 3D vector of relative integers $(k_1,k_2,k_3)$)
\begin{equation}
{\partial_t { \hat v}_\alpha(t,\vk)}  =  -\frac{i} {2} {\mathcal P}_{\alpha \beta \gamma}(\vk) \sum_{\bf p} {\hat v}_\beta(t,{\bf p}) {\hat v}_\gamma(t,{\bf k-p})
-\nu k_\beta k_\beta {\hat v}_\alpha(t,\vk) +i \sqrt{D } \epsilon_{\alpha\beta\gamma}k_\beta  {\hat f}_\gamma
\label{eq_discrt}
\end{equation}
where ${\mathcal P}_{\alpha \beta \gamma}(\vk)=k_\beta P^\perp_{\alpha \gamma}(\vk)+k_\gamma P^\perp_{\alpha \beta}(\vk)$  and the convolution in (\ref{eq_discrt}) is truncated (set to zero) for $k_\alpha k_\alpha \ge  k_{\rm max}^2$.

When $D =\nu=0$, this truncated Euler equation is a time-reversible system which exactly conserves the kinetic energy
$E=\sum_{k}E(t,k)$, where the energy spectrum $E(t,k)$ is
defined by averaging ${\hat \vv}(t,{\vk'})$ on spherical shells of width $\Delta k = 1$,
\begin{equation}
E(t,k) = {\frac1 2} \sum_{k-\Delta k/2< |{\vk'}| <  k + \Delta k/2} |{\hat \vv}({\vk'},t)|^2 \, .
\label{eq_energy}
\end{equation}
At large times, the truncated 3D Euler equations is known to reach an equilibrium, which is a statistically stationary exact Gaussian solution, with energy spectrum $E(k)= c k^2$~\cite{Cichowlas2005}. When $\nu$ and $D$ are non-zero, identical statistically stationary solutions also exist. A realisation of such a solution, with $u_{rms}=\sqrt{2 E}=\frac{1}{2}$, is used as initial data to start our numerical simulations. When $\nu$ is non-zero, $D$ is adjusted so that the corresponding stationary solution also has $u_{rms}=\frac{1}{2}$ (see~\cite{Brachet2022} for details on a similar setting for the 1D Burgers equation).

We use fields with Taylor-Green symmetries (see Ref.~\cite{Cameron2017} for details). Time-stepping of the deterministic terms of~\eqref{eq_discrt} is performed using a $4^{\rm th}$-order Runge-Kutta method and the random thermal noise term is added independently.
Two-dimensional slices of the 3D computed Fourier-space velocity field are output periodically every $\delta t$ to files.
Post-processing the files, time-correlation functions and spatio-temporal power spectra up to the Nyquist pulsation $\omega_{\rm max}=\pi/\delta t$ are estimated using standard algorithms~\cite{Press1997}.
The runs presented here use spatial resolution of $256^3$ and $u_{rms}=\frac{1}{2}$.The inviscid run uses the output frequency $\delta t=0.1$ and a total integration time of  $t=1024$. The viscous run with $\nu=0.016$ uses $\delta t=0.0125$ and a total integration time of  $t=128$.

\subsection{Scalings of correlation times}

\begin{figure}
\includegraphics[width=0.5\linewidth]{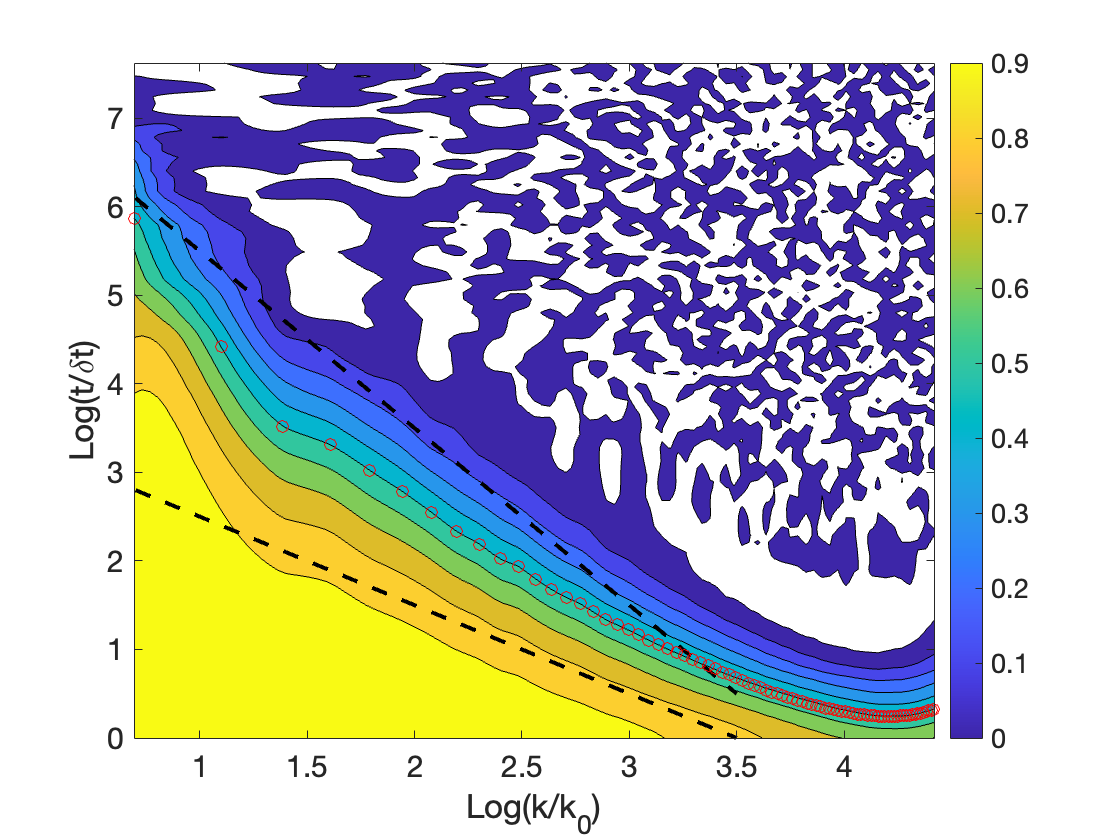}
\includegraphics[width=0.5\linewidth]{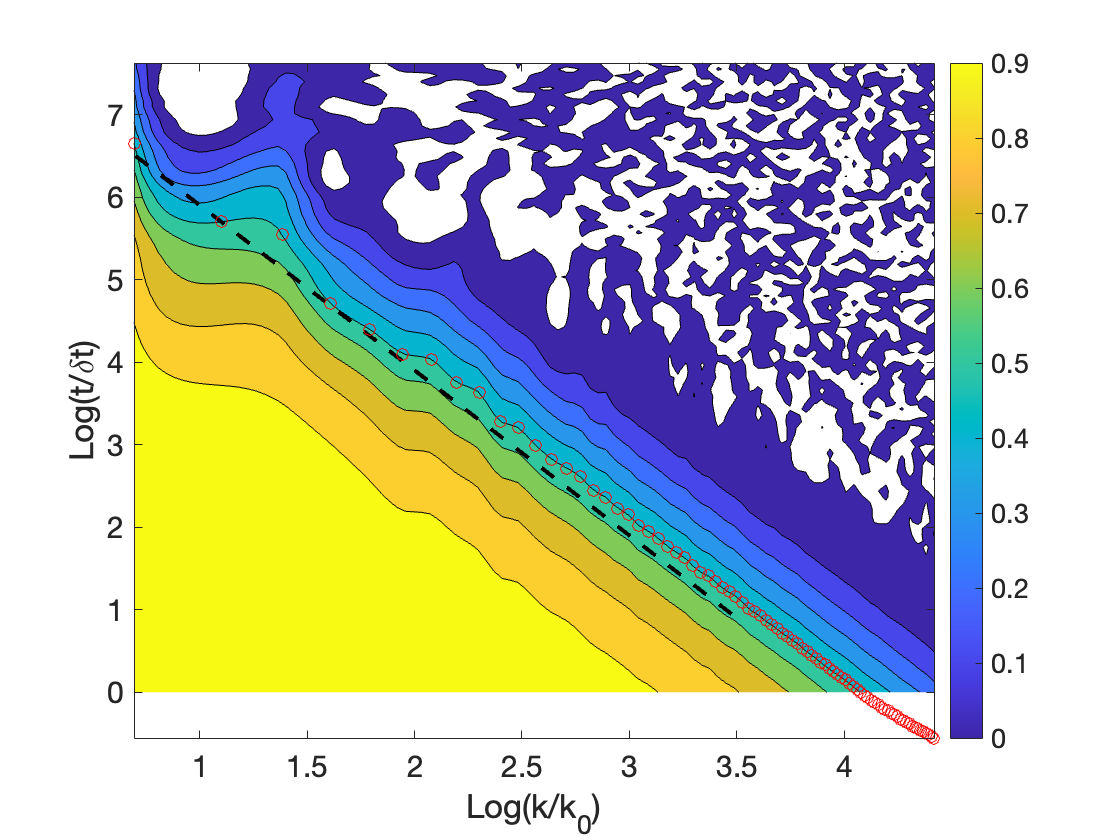}
\caption{
Contour plots of the correlation function $\bar{C}(t,k)/\bar{C}(0,k)$ represented in the natural logarithm $(\log(k),\log(t))$ plane;
(left):  inviscid scaling obtained at
$\nu=0$; (right):  viscous scaling obtained at $\nu= 0.016$. $\bar{C}(t,k)/\bar{C}(0,k)$  contour levels are drawn from $0.0$ to $0.9$ and spaced by $0.1$.
The dashed black lines indicate (left) $t\sim k^{-1}$ and $t\sim k^{-2}$; (right) $t\sim k^{-2}$. 
The points $\tau_{1/2} (k)$, computed independently using $\bar{C}(\tau_{1/2},k)={\frac{1}{2}}\bar{C}(0,k)$
are indicated by red circles.
}
\label{fig:contours1}
\end{figure}

We first study the correlation function $\bar{C}(t,k)$, and the related half decay time $\tau_{1/2} (k)$.
 In order to visualise its scaling behaviour, we represent in Fig.~\ref{fig:contours1} the map of the normalised correlations
$\bar{C}(t,k)/\bar{C}(0,k)$ in the $(\log(t),\log(k))$ plane.
Indeed, scaling in this representation simply corresponds to equal values of $\bar{C}(t,k)/\bar{C}(0,k)$ along straight lines.
\begin{figure}
\includegraphics[width=0.5\linewidth]{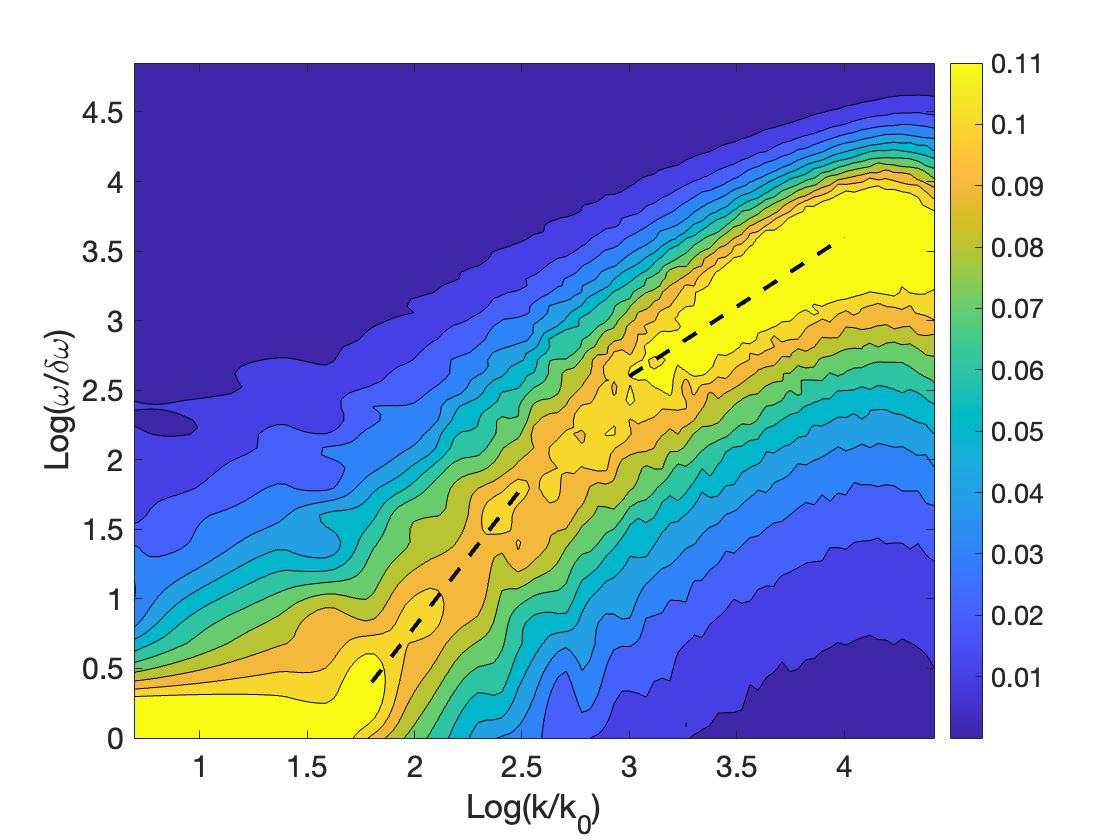}
\includegraphics[width=0.5\linewidth]{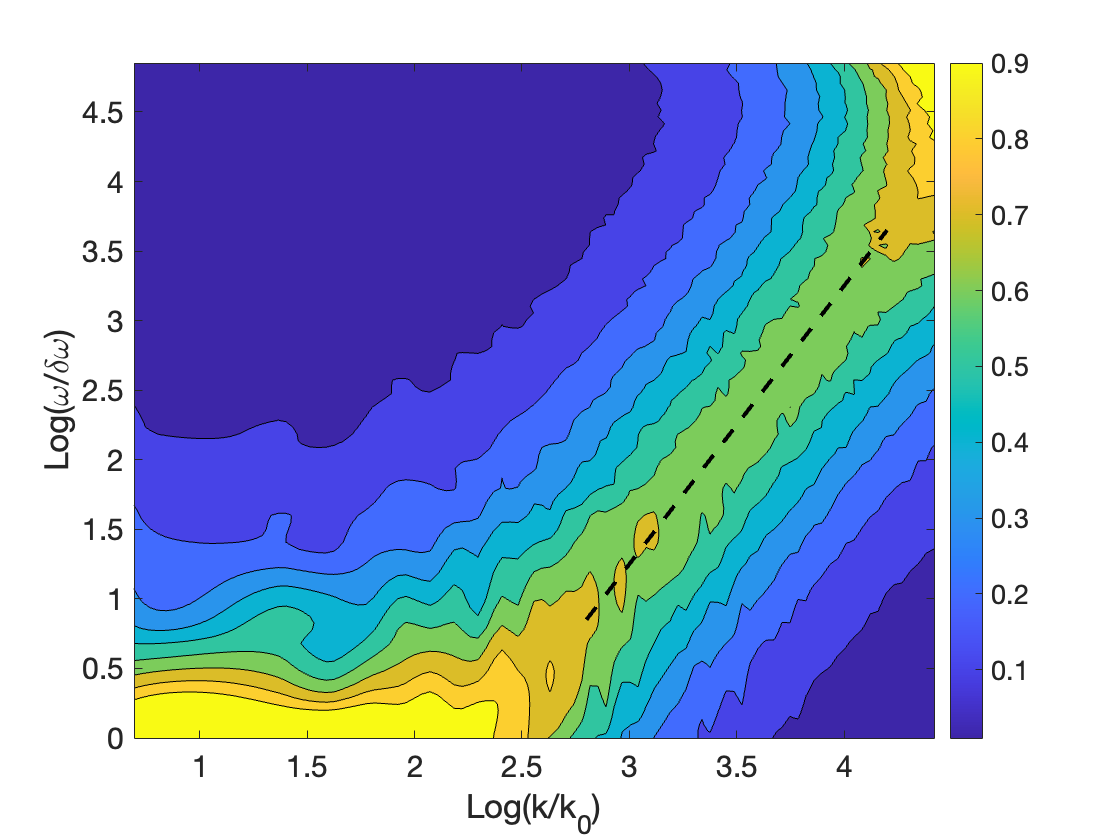}
\caption{
Contour plots of $\omega S(\omega,k)$, where $S(\omega,k)$ is the power spectrum, {\it i.e} the Fourier transform of $\bar{C}(t,k)/\bar{C}(0,k)$, represented in the natural logarithm $(\log(k),\log(\omega))$ plane; (left): inviscid scaling obtained at
$\nu=0$; (right): viscous scaling obtained at $\nu= 0.016$.
The dashed black lines indicate (left) $\omega\sim k^{1}$ and $\omega\sim k^{2}$; (right) $\omega\sim k^{2}$. 
 The  contour levels of  $\omega S(k,\omega)$ are: (left) from $0.0$ to $0.09$ (spaced by $0.01$) and  (right) $0.0$ to $0.9$ (spaced by $0.1$).
}
\label{fig:contours2}
\end{figure}
The red circles \footnote{Red circles standing right on top of the $0.5$ contour line validates the interpolation scheme used to obtain the logarithmic representation from the raw numerical data that is equally spaced both in time and wavenumber.} indicate the points $\tau_{1/2} (k)$
obtained independently by solving for each value of $k$ the implicit definition $\bar{C}(\tau_{1/2}(k),k)={\frac{1}{2}}\bar{C}(0,k)$.
The left panel of Fig.~\ref{fig:contours1} corresponds to  the inviscid case $\nu=D =0$.  The black dashed lines indicate the $t\sim k^{-1}$ and $t\sim k^{-2}$ scaling laws, which are both observable.
 In fact, the small-time scaling behaviour
 of the correlation with timescale $t\sim k^{-1}$ was already predicted in the inviscid limit in Ref.~\cite{Cichowlas2005,Cameron2017}, in agreement with the FRG exact result for small time delays~\eqref{eq:solutionC}. This scaling is apparent on the  level lines from 0.9 to 0.5  which closely follow this law. Interestingly, at larger times, another scaling emerges, indicated by the level line $0.1$ which is definitely steeper and nearer to a $t\sim k^{-2}$ behaviour. This is an evidence for the FRG prediction for the large time-delay behaviour~\eqref{eq:solutionC}.
On the right panel  of Fig.~\ref{fig:contours1}, the results for the stochastic  Navier-Stokes equation obtained at $\nu= 0.016$ is displayed.
The black dashed line indicates the theoretical scaling $t\sim k^{-2}$, corresponding to the diffusive FNS behaviour, which  is well-followed for this large viscosity.

The scaling behaviour can also be observed on the
  spectrum $S(k,\omega)$, defined as the Fourier transform in time of $\bar{C}(t,k)/\bar{C}(0,k)$. The numerical noise on this quantity can be reduced
 by considering  $\omega S(k,\omega)$ instead~\cite{Gottlieb1977,Press1997}. Fig.~\ref{fig:contours2} shows the contour plot of $\omega S(k,\omega)$.
 Scaling laws again correspond to straight lines in the $(\log(\omega),\log(k))$ plane. 
The left panel of Fig.~\ref{fig:contours2} displays the inviscid  scalings, where
the black dashed lines indicate $\omega \sim k$ and $\omega \sim k^{2}$ scaling laws. It is apparent that both scalings are present: $\omega \sim k^{2}$ for small $\omega$ and $\omega \sim k$ for large  $\omega$.
On the right panel, the viscous scaling obtained at $\nu= 0.016$ is displayed.
The black dashed line indicates the expected theoretical $\omega\sim k^{2}$ diffusive law, which is again well followed.

\begin{figure}
\includegraphics[width=0.5\linewidth]{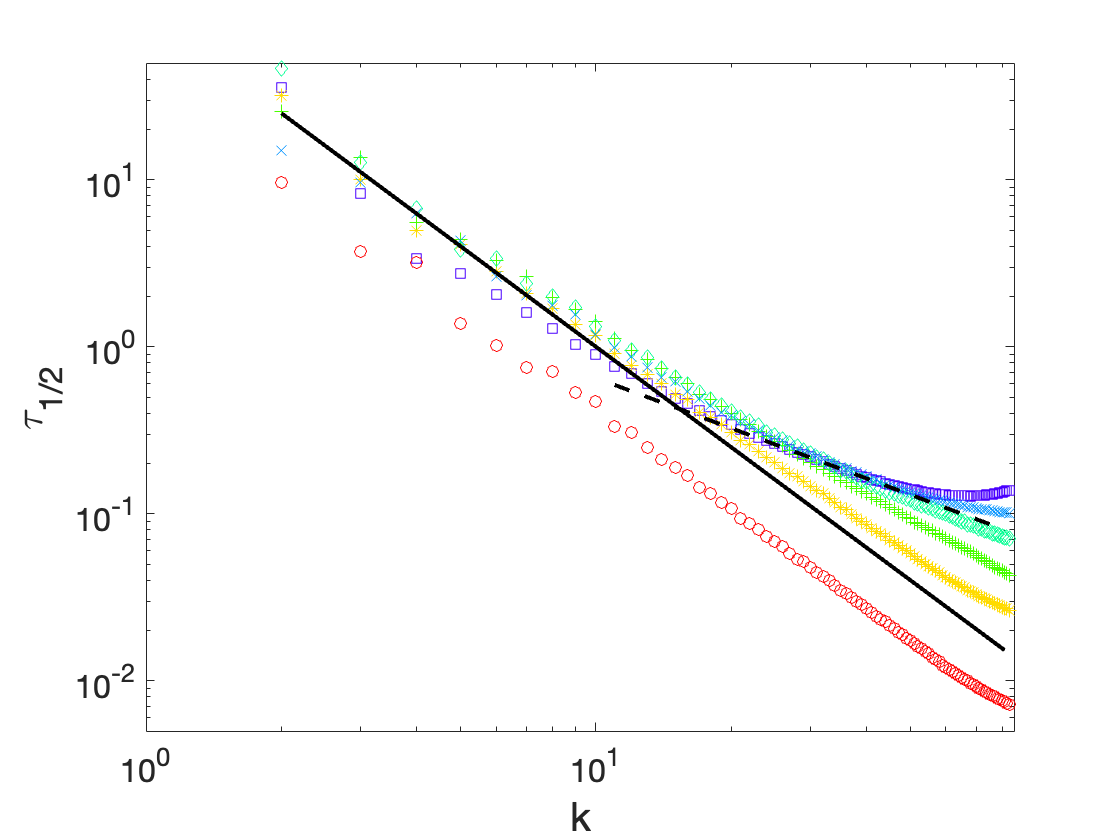}
\includegraphics[width=0.5\linewidth]{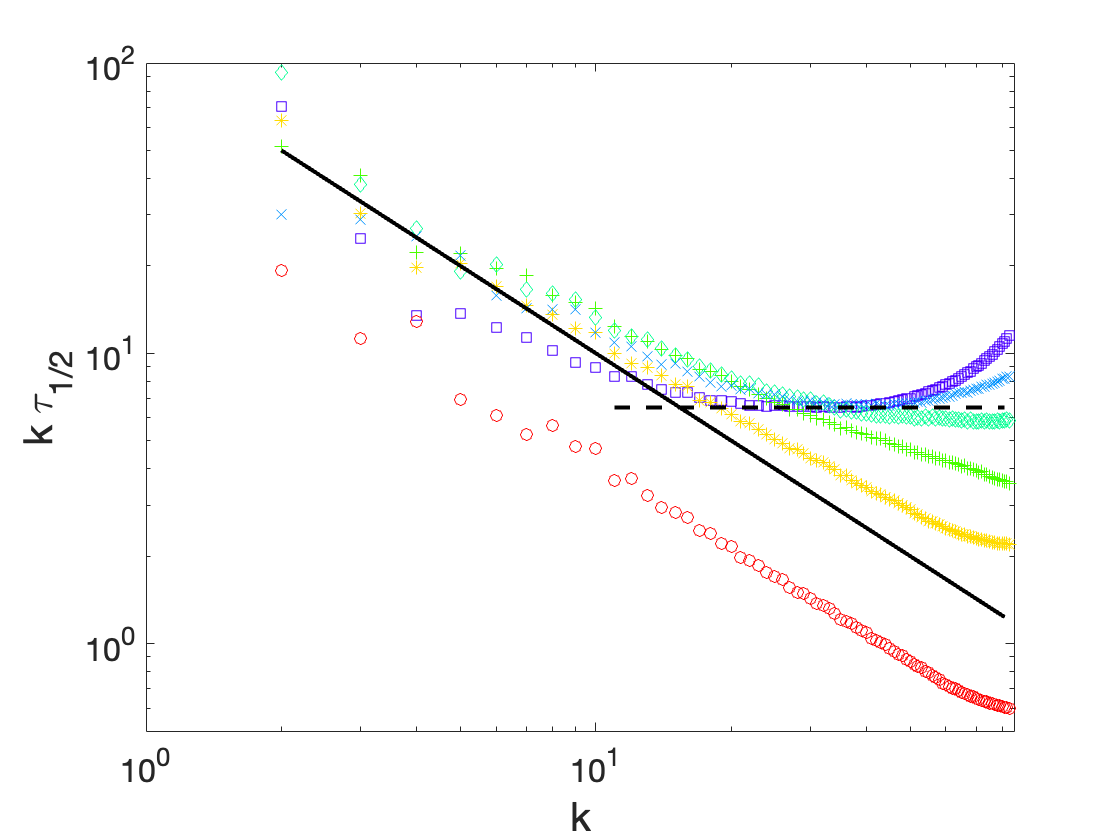}
\caption{Left panel: half decay time $\tau_{1/2}$  of the correlations versus wavenumber $k$ for different viscosities obtained from numerical simulations. At the largest viscosity, only the viscous  scaling $\tau\sim k^{-2}$ is visible, indicated by the solid line. At the smallest viscosities, the behaviour at large wavenumbers shows the inviscid scaling $\tau\sim k^{-1}$, indicated by the dashed line.
Right panel:  same as in left panel, but compensated by $k$,
 {\it i.e} $k \, \tau_{1/2}(k)$ versus $k$.
Markers correspond on both panels to:
$\nu= 0.016$: $o$,  $\nu=0.0044$: asterisk,  $\nu=0.0022$: $+$, $\nu=0.0011$: diamond, $\nu=0.00055$: x  and $\nu=0$: squares  (colors from red to blue).
}\label{fig:tau-one-half} 
\end{figure}

One can study as well the emergence of the $z=1$ scaling varying the viscosity. As highlighted in Fig.~\ref{fig:omega-half} obtained with FRG, one expects that at finite but non-zero viscosity, this scaling also appears at large wavenumbers, all the more pronounced that the viscosity is small. We show on the left panel of Fig.~\ref{fig:tau-one-half}  the half decay time $\tau_{1/2}(k)$  obtained from numerical simulations from $\nu=0.016$ to $\nu=0$, and on the right panel the same data compensated by $k$, {\it i.e} $k\tau_{1/2}(k)$, such that the $z=1$  scaling appears in this representation as a plateau.  Note that at the highest momenta, all the curves saturate due to truncation effects. For the largest viscosity, only the diffusive regime $z=2$ is observed, while as the viscosity decreases, the behaviour of the large momentum mode bends towards a $z=1$ scaling. The plateau is visible on the right panel for viscosities $\nu \lesssim 0.0022$.  The numerical results are thus in perfect accordance with the FRG prediction.

\section{Relevance of the IFNS scaling in physical systems}
\label{sec:exp}

An important question is whether the unveiled scaling behaviour is observable in a real fluid at equilibrium. In Fig.~\ref{fig:omega-half}, the momentum $p$ is measured in units of $\Lambda$  and the  viscosity $\nu$ in units of $(\Lambda\chi)^{1/2}$. The scale $\Lambda^{-1}$ represents the length scale at which the hydrodynamical description is valid, it should be chosen large enough with respect to the typical microscopic length, for instance $\Lambda^{-1}\sim 10 \lambda­_{\text{mfp}}$, where $\lambda­_{\text{mfp}}$ is the mean free path. This would set for air the units to be $1.5\times10^4 \text{cm}^{-1}$ for momentum and 0.0007 cm$^2$/s for viscosity, and for water $4\times10^6 \text{cm}^{-1}$ for momentum and 0.0004 cm$^2$/s for viscosity at $T=300$K.
 To observe the $z=1$ scaling, we need to set $\bar g_\Lambda \gg 1$, or $\nu \ll 1 \times 10^{-4}\text{cm}^2$/s, and this behaviour emerges for  wavenumbers starting from $10^4 \text{cm}^{-1}$ for air and $10^5 \text{cm}^{-1}$ for water. However,  even without substituting specific constants, the estimation of the viscosity for these fluids ({\it eg.} in Ref.~\cite{Bandak2021}, Eq.~(2.26), which agrees with numerical calculation of viscosity at the scale of the mean free path in Ref.~\cite{Donev2014}) shows that $\bar g_\Lambda$ remains small for molecular fluids as long as the scale  $\Lambda^{-1}$ is larger than the mean free path. Thus, the $z=1$ scaling probably does not lie within the physical region for molecular fluids at equilibrium. On the other hand, it becomes observable when the fluid is turbulent.

\section{Conclusion}\label{sec:conclusion}

In this work, we have investigated the stochastic 3D Navier-Stokes equation in the presence of thermal noise, which corresponds to model A of FNS, using both direct numerical simulations and functional renormalisation group.
The aim was to explore what happens when taking the inviscid limit in this model,  which had been studied so far only in the perturbative regime, corresponding to large viscosities.
 We have indeed identified a new non-perturbative, strong-coupling, fixed point, called IFNS, found in the inviscid (Euler) limit of model A, and which also controls, when the viscosity is finite, the large momentum behaviour of the correlation function. This fixed point is characterised by a critical  dynamical exponent $z=1$ at small time delays.
Moreover, the FRG result shows that the behaviour of the temporal correlations crosses over at large time delays to a much slower decay with $z=2$. This second time regime had remained elusive in direct numerical simulations so far due to the difficulty to properly resolve  long time delays. Our numerical simulations of model A have allowed us to observe both regimes, yielding  the first numerical evidence of the long-time regime for 3D Navier-Stokes equation. The simulations have thus fully confirmed the FRG predictions.
This work has brought a further theoretical insight in the behaviour of the stochastic Navier-Stokes equation, although the IFNS scaling may not be observable in a real fluid at equilibrium.

Remarkably, the unveiled $z=1$ scaling is common to the inviscid regime of both the Burgers and Navier-Stokes equations. Moreover, the value of $z$ is  independent of the dimension, as well as of the precise correlations of the stochastic term $\vf$ provided it has a Gaussian distribution~\cite{Tarpin2018,Gosteva2024}. Thus it holds both for equilibrium situations (when $\vf$ is a thermal noise, and the stationary state is a fluid at rest) and non-equilibrium situations (when $\vf$ is a large-scale forcing, and the stationary state is a turbulent fluid), even though the corresponding static behaviours, {\it eg.} the kinetic energy spectra, are completely different.
We have provided the explanation of this robustness, as we have shown that the large momentum behaviour of correlation functions is completely fixed and solely determined by the symmetries. Indeed, all these hydrodynamical equations share the same fundamental (extended) symmetries: the time-dependent Galilean symmetry and time-dependent shifts of the response fields (although the latter bears slightly different forms in Burgers and NS equations). These symmetries, through the related Ward identities, allow for an exact closure of the FRG flow equations in the asymptotic regime of large momenta. Their solution endows identical forms for Burgers or NS flows, differing only by non-universal constants, and featuring the exact $z=1$ exponent. The FRG has thus proved very efficient to understand the temporal behaviour of correlation functions. In perspective, it would be valuable to use it to gain more understanding of static properties,  encoded for instance in  structure functions, and in particular of intermittency effects.

\vskip6pt

\ack{LC wishes to thank G. Eyink for suggesting this work and for his valuable comments about the physical relevance of the IFNS scaling. LG acknowledges support by the MSCA Cofund QuanG (Grant Number : 101081458) funded by the European Union. Views and opinions expressed are however those of the authors only and do not necessarily reflect those of the European Union or Université Grenoble Alpes. Neither the European Union nor the granting authority can be held responsible for them.}

%%%%%%%%%% Insert bibliography here %%%%%%%%%%%%%%

%\begin{thebibliography}{9}

%\bibliography{../Biblio/bibBurgers,../Biblio/bibNPRG,../Biblio/bibNS,../Biblio/bibKPZ}
%\bibliographystyle{prsty-title}

\end{document}